\documentclass[useAMS,usenatbib,usegraphicx]{mn2e}
\usepackage{bm}
\usepackage{times}

\title[Dust content probed by gamma-ray burst afterglows]
{Evolution of dust content in galaxies probed by gamma-ray burst
afterglows}
\author[Kuo et al.]{Tzu-Ming Kuo,$^{1,2}$\thanks{E-mail:
    Raphaecaro@gmail.com}
Hiroyuki Hirashita,$^1$ and
Tayyaba Zafar$^{3,4}$\\ 
$^1$Institute of Astronomy and Astrophysics, Academia Sinica,
P.O. Box 23-141, Taipei 10617, Taiwan\\
$^2$Department of Physics, National Taiwan University,
Taipei 10617, Taiwan\\
$^3$Aix Marseille Universit\'{e}, CNRS, LAM (Laboratoire
d'Astrophysique de Marseille) UMR 7326, 13388, Marseille, France\\
$^4$Department of Physics, University of the Punjab, Quaid-i-Azam
Campus, Lahore-54590, Pakistan
}
\date{2013 August 29}

\pagerange{\pageref{firstpage}--\pageref{lastpage}} \pubyear{2013}

\begin{document}
\label{firstpage}
\maketitle

\begin{abstract}
Because of  {their} brightness, gamma-ray burst (GRB) afterglows are viable
targets  {for investigating} the dust content in their host galaxies.
Simple intrinsic
spectral shapes of GRB afterglows allow us to derive the dust extinction.
Recently, the extinction data of GRB afterglows
are compiled up to redshift $z=6.3$, in combination with
hydrogen column densities and metallicities. This data set enables us
to investigate the relation between dust-to-gas ratio and metallicity  {out}
to high redshift  {for} a wide metallicity range. By applying our evolution
models of dust content in
galaxies, we find that the dust-to-gas ratio derived from GRB afterglow
extinction data are excessively high such that they can be explained
with a fraction of gas-phase metals condensed into dust
($f_\mathrm{in}$) $\sim 1$, while theoretical calculations on dust formation
in the wind of asymptotic giant branch stars and in the ejecta of
Type II supernovae suggest a much more moderate condensation
efficiency ($f_\mathrm{in}\sim 0.1$). Efficient dust growth in dense clouds
has difficulty in
explaining the excessive dust-to-gas ratio at metallicities
$Z/\mathrm{Z}_{\sun}<\epsilon$, where $\epsilon$ is the star formation
efficiency of the dense clouds. However, if $\epsilon$ is as small
as 0.01, the dust-to-gas ratio at $Z\sim 10^{-2}$ Z$_{\sun}$ can
be explained with $n_\mathrm{H}\ga 10^6$ cm$^{-3}$.
Therefore, a dense environment hosting dust
growth is required to explain the large fraction of metals condensed into dust,
but such clouds should have low star formation efficiencies to avoid
rapid metal enrichment by stars.
\end{abstract}


\section{Introduction}
\label{intro}

One of the important problems in astrophysics is the origin and
evolution of dust in the Universe, since various aspects of
galaxy evolution are significantly influenced by
the optical and material properties and the total abundance
of dust. For example, dust governs the absorption, scattering,
and reemission of the stellar light, affecting
the radiative transfer in the interstellar
medium (ISM) \citep[e.g.][]{yajima12}. Furthermore, the surface of dust
grains is the main site for the formation of  some molecular species,
especially H$_2$, which could
affect the star formation properties of galaxies \citep{hirashita02,yamasawa11}.
Therefore, clarifying the origin and evolution of dust content is essential for revealing
how galaxies have evolved in the Universe.

It is widely believed that the scenario of the evolution of dust
content in galaxies comprises dust formation in stellar ejecta,
dust destruction in supernovae (SN) remnants, and grain growth
by the accretion of metals onto preexisting grains in molecular
clouds
\citep[e.g.][]{dwek98,hirashita99,inoue03,zhukovska08,valiante11,mattsson12}.
These processes depend on the age and metallicity of galaxies.
In particular, the dominant mechanism of dust enrichment is
suggested to switch from the supply by the stellar ejecta to the
accretion of metals at a certain metallicity level
\citep{inoue11,asano13}.

For the purpose of acquiring the general trend of the evolution
of dust content in galaxies at different ages and metallicities,
the approaches using extinctions of bright sources of which
the intrinsic spectra are well known, for example quasars
(QSOs) and GRB afterglows, are regarded as viable methods,
since they are bright enough to be detected even at high redshift.
Quasars are usually used to probe the foreground galaxies
in absorption, while GRB afterglows are often utilized to probe the
ISM of their own host galaxies.

Recently, \citet{zafar13} compiled and analyzed GRB afterglow
data in a wide redshift range of $z=0.1-6.3$. By using the
$A_V/N_\mathrm{H}$ [$A_V$ is the extinction at the $V$ band
 {(0.55 $\micron$)}, and $N_\mathrm{H}$ is the H \textsc{i}
column density] ratio
as an indicator of dust-to-gas ratio, they show that the relation
between dust-to-gas ratio and metallicity is on a natural
extension of the local galaxy sample, even at low
metallicities down to $Z\sim 10^{-2}$ Z$_{\sun}$
($Z$ is the metallicity and Z$_{\sun}$ is the solar metallicity).
This indicates that the fraction of metals condensed into dust
is as high as the local
galaxy sample even at such a low metallicity.
They did not find any systematic difference
between the GRB sample and a QSO absorption sample
used as a comparison sample,
rejecting the systematics of the GRB sample relative to other
samples. Therefore, they argue that there is a close correspondence
between dust formation and metal formation; in other words,
any delay between the formation of metals and dust must be
shorter than typically a few Myr [i.e.\ the time-scale of metal
enrichment by supernovae (SNe)]. They finally propose two possibilities of
dominant dust formation mechanisms consistent with the
close association between metals and dust: (i) rapid dust enrichment
by condensation in the ejecta of SNe; and (ii) rapid grain growth by the
accretion of gas-phase metals onto dust grains in the ISM.

In this study, we utilize the GRB afterglow extinction data
to investigate the evolution of dust content in their host galaxies.
In particular, we judge if the above two possibilities (i) and (ii)
are theoretically supported or not, by applying a dust enrichment
model developed in our previous studies. Through this work,
we will be able to obtain or constrain some essential parameters
for dust enrichment, especially, the efficiencies of
dust condensation and growth.

This paper is organized as follows. In Section \ref{sec:data},
we present the observational data adopted.
In Section \ref{sec:model}, we overview our theoretical models
used to interpret the data. In Sections \ref{sec:result}
and \ref{sec:discussion}, we
provide results and discussions, respectively. The conclusions
are given in Section \ref{sec:conclusion}. In this paper,
we adopt $\mathrm{Z}_{\sun}=0.02$ for the solar metallicity.


\section{Data}\label{sec:data}

\subsection{Extinction data of GRB afterglows}

We adopt the extinction data of a sample of GRB afterglows (simply called 
GRBs hereafter) from \citet{zafar13}. They collected data from the literature 
for GRBs on the basis that they have: $i)$ optical extinction estimates 
(derived from the X-ray-to-optical/near-infrared spectral energy distribution 
fitting; see \citealt{zafar11, zafar12}, for the method), $ii)$ Zn or S based 
metal column densities (since they are rarely condensed into dust), and 
$iii)$ {H}\,{\sc i} column density (wherever possible) measurements. 
In total, they compiled 25 GRBs with such available measurements. 
The models are mainly constrained by the objects whose metallicity,
$N_\mathrm{H}$, and $A_V$ are all detected
($A_V$ is
the rest-frame $V$-band extinction 
and $N_\mathrm{H}$ is the {H}\,{\sc i} column density of the GRB);
that is, if only an upper
limit is obtained for either of those values, the data is excluded, and
we are left with 9 GRBs.

\citet{zafar13} compared the metals-to-dust ratios of GRBs, QSO foreground 
damped Ly$\alpha$ systems (DLAs), and galaxy-lensed QSOs to the value 
obtained in the Milky Way (MW), the Large Magellanic Cloud (LMC), and 
the Small Magellanic Cloud (SMC). They did not find any systematic 
difference between the GRB, QSO-DLA, and lensed galaxies samples, 
rejecting the systematics of the GRB sample relative to other samples.
Therefore, for the uniformity of the sample, we concentrate on the GRB 
sample since inclusion of the QSO-DLA and galaxy-lensed QSO samples 
do not affect our conclusion. We refer to \citet{vladilo04} for
a detailed analysis of the dust evolution in DLAs.

In this paper, we focus on the relation between dust-to-gas mass ratio 
$\mathcal{D}$ and metallicity $Z$ (called $\mathcal{D}-Z$ relation).
To convert $A_V/N_\mathrm{H}$, which is used as an indicator of 
dust-to-gas ratio by \citet{zafar13}, to $\mathcal{D}$, we apply
the formula explained  in the next subsection.
The information about extinction curve of each individual GRB is
collected from the references provided in Table~1 of
\citet{zafar13}.

\subsection{Dust-to-gas ratio}\label{subsec:dg}

In order to obtain the dust-to-gas ratio $\mathcal{D}$ from
the extinction data in GRBs, we refer to the approaches established
by \citet{pei92}. In this paper, we only consider H \textsc{i} for
the hydrogen
content, and neglect the contribution from
molecular hydrogen H$_2$ and ionized hydrogen H \textsc{ii}.
Ionized hydrogen is unlikely to contribute largely to the total
hydrogen mass \citep{spitzer78}, while H$_2$ may have a
large contribution. However, in nearby low-metallicity
dwarf galaxies, molecular gas traced by CO is rarely detected, although
it is not clear whether it is due to a real lack of H$_2$ or a
different conversion factor (C may be in the form of C \textsc{ii}
rather than CO; e.g.\ \citealt{madden00}). Because of such an uncertainty in
H$_2$, we neglect the contribution from H$_2$.
Below we briefly review the conversion formula from
$A_V/N_\mathrm{H}$ to $\mathcal{D}$. See \citet{pei92} for
details.

The dust-to-hydrogen mass ratio is proportional to the
extinction optical depth divided by $N_\mathrm{H}$
(in units of cm$^{-2}$) as
\begin{eqnarray}
\rho_\mathrm{dust}/\rho_\mathrm{H}\equiv\chi\cdot
10^{21}(\tau_B/N_\mathrm{H}) ,
\label{eq:chi}
\end{eqnarray}
where $\rho_\mathrm{dust}$ and $\rho_\mathrm{H}$ are
the mass densities of dust and hydrogen, respectively,
$\chi$ is a constant depending both on the
optical and material properties of dust grains, and
$\tau_B$ is the extinction optical depth at the $B$-band
 {(0.44 $\micron$)}.
Using equation (\ref{eq:chi})
and converting $\tau_\lambda$ (extinction optical depth
at wavelength $\lambda$)
into $A_\lambda$ (extinction at $\lambda$ in units of magnitude),
we obtain the following estimate for the dust-to-gas mass ratio,
$\mathcal{D}$:
\begin{eqnarray}
\mathcal{D}& \hspace{-2mm}\equiv & \hspace{-2mm}
\frac{\rho_\mathrm{dust}}{\rho_\mathrm{gas}}
=\frac{10^{21}}{1.4}\frac{\ln 10}{2.5}
\left(\frac{1+R_V}{R_V}\right)\chi\left(\frac{A_V}{N_\mathrm{H}}\right)
\equiv\Lambda\left(\frac{A_V}{N_\mathrm{H}}\right) ,\nonumber\\
\end{eqnarray}
where $\rho_\mathrm{gas}=1.4\rho_\mathrm{H}$ (the factor 1.4
comes from the correction for elements other than hydrogen),
$R_V\equiv A_V/(A_B-A_V)$ is the ratio of total-to-selective
extinction, and $\Lambda$ is
referred to as the converting factor. By using the numerical values
in \citet{pei92}, we obtain
$\Lambda =1.13\times 10^{19}$, $1.56\times 10^{19}$,
and $2.03\times 10^{19}$  cm$^{-2}$ mag$^{-1}$ for
the MW, LMC, and SMC extinction curves, respectively.
We choose one of these three values for each object
according to the extinction curve adopted by
\citet{zafar13} (see their Table 1).


\subsection{Nearby galaxy data}\label{subsec:nearby}

Since the dust evolution models have often been `calibrated'
with nearby galaxy data \citep[e.g.][]{lisenfeld98,dwek98,hirashita99},
it would be interesting to examine if there is a difference
in the $\mathcal{D}-Z$ relation between the GRB sample
and nearby galaxies.
For the uniformity of data, we select the samples
compiled by \citet{hirashita11} (see the references therein
for original data), who construct the sample based on an
\textit{AKARI} sample: eight
blue compact dwarf galaxies (BCDs) and three spiral galaxies.
Note that the dust content is measured by far-infrared
emission, not by extinction. We also include two spiral galaxies
whose dust content is measured by extinction to confirm that
there is no systematic difference in the estimate of dust
content. More data are seen in e.g.\ \citet{engelbracht08}:
the addition of such a data set
only makes the plots dense without changing the trend
in the $\mathcal{D}-Z$ relation.

\section{Models}\label{sec:model}

\subsection{Dust enrichment}

For the purpose of investigating the evolution of  {dust} content in
a galaxy, a chemical enrichment model which describes
the time evolution of gas, metals, and dust in a galaxy is adopted.
We apply a simple model used by \citet{hirashita11}.
The simplicity of the model is the advantage for our purpose of
seeing the response of the $\mathcal{D}-Z$ relation to
dust formation and
destruction processes. We suppose the galaxy to be a closed box;
that is, we neglect inflow and outflow. The effects of inflow and
outflow  {have} only a minor effect on the $\mathcal{D}-Z$
relation \citep{asano13}. We also assume that the mixing
of substances in galaxies is immediate and complete such that the
system is treated as a one-zone environment. To make the
problem analytically tractable (without losing the essence), we adopt an
instantaneous recycling approximation, in which we assume that
stars whose lifetimes are shorter than 5 Gyr return the gas soon
after their formation. This means that we do not divide the
metal and dust production by SNe and AGB stars for simplicity.
Instead, we separately discuss the contributions from SNe and AGB
stars in details when we interpret the results
(Section \ref{subsec:condensation}). Similar kinds of analytical
models are also adopted and successfully catch the
essential features in the relation between dust-to-gas ratio
and metallicity
\citep{lisenfeld98,hirashita99,inoue03,inoue11,mattsson12,mattsson13}.
For the analytic solutions and asymptotic behaviours of
the models, we refer to \citet{inoue11} and \citet{mattsson13}.

The model equations are finally reduced to the relation between
$\mathcal{D}$ and $Z$ \citep{hirashita11},
since the abundances of metals and dust are tightly connected:
\begin{eqnarray}
\mathcal{Y}\frac{\mathrm{d}\mathcal{D}}{\mathrm{d}Z}=f_\mathrm{in}
(\mathcal{R}Z+\mathcal{Y})-(\beta_\mathrm{SN}+\mathcal{R})
\mathcal{D}+\frac{1}{\psi}\left[
\frac{\mathrm{d}M_\mathrm{dust}}{\mathrm{d}t}\right]_\mathrm{acc},
\label{eq:dDdZ}
\end{eqnarray}
where $\mathcal{R}$ is  the returned fraction of the mass from formed stars,
$\mathcal{Y}$ is the mass fraction of metals that is newly produced and
ejected by stars, $f_\mathrm{in}$ is the condensation efficiency of
metals into dust in stellar ejecta ($f_\mathrm{in}=1$ means that
all the metals are condensed into dust), $\psi$ is the
star formation rate,
$[\mathrm{d}M_\mathrm{dust}/\mathrm{d}t]_\mathrm{acc}$ is
the rate of increase of dust mass by the accretion of metals onto
preexisting grains (this term is formulated separately in
Section \ref{subsec:growth}), and $\beta_\mathrm{SN}$ is
the efficiency of destruction of dust by shocks in SN remnants,
which is defined by
$\beta_\mathrm{SN} = \epsilon_\mathrm{s}M_\mathrm{s}\gamma /\psi$,
with $\epsilon_\mathrm{s}$ being the fraction of dust destroyed in
a single SN blast, $M_\mathrm{s}$ being the gas mass swept
per SN blast, and $\gamma$ being the SN rate, as introduced by
\citet{mckee89}. Following \citet{hirashita11}, we adopt
$\mathcal{R}=0.25$ and $\mathcal{Y}=0.013$ throughout this paper,
and $\beta_\mathrm{SN}=9.65$ unless otherwise stated.

\subsection{Grain growth by the accretion of metals onto preexisting grains}
\label{subsec:growth}

We adopt the formulation developed by \citet{hirashita11} for
the increasing rate of dust mass by the accretion of metals onto
preexisting grains (called grain growth hereafter):
\begin{eqnarray}
\left[\frac{\mathrm{d}M_\mathrm{dust}}{\mathrm{d}t}\right]_\mathrm{acc}
=\frac{\beta\mathcal{D}\psi}{\epsilon},\label{eq:dmdt_acc}
\end{eqnarray}
where $\epsilon$ is the star formation efficiency in molecular clouds
(assumed to be 0.1 unless otherwise stated), and
$\beta$ is the increment of dust mass in molecular clouds, which can
be estimated as
\begin{eqnarray}
\beta\simeq\left[
\frac{\langle a^3\rangle_0}{3y\langle a^2\rangle_0+3y^2\langle a\rangle_0
+y^3}
+\frac{1-\xi }{\xi}\right]^{-1},\label{eq:beta}
\end{eqnarray}
where $\xi\equiv 1-\mathcal{D}/Z$ is the fraction of metals in gas phase,
$\langle a^\ell\rangle_0$ is the $\ell$th moment of grain radius
(we adopt $\langle a\rangle=1.67\times 10^{-3}~\micron$,
$\langle a^2\rangle=4.68\times 10^{-6}~\micron^2$, and
$\langle a^3\rangle=7.41\times 10^{-8}~\micron^3$ based on
a grain size distribution with power index $-3.5$ and  {lower and upper}
limits for the grain radius 0.001 and 0.25 $\micron$, respectively;
\citealt{mathis77}) and $y\equiv a_0\xi\tau_\mathrm{cl}/\tau$,
with $a_0 = 0.1~\micron$
being  {an} arbitrarily given typical radius of dust grains, $\tau_\mathrm{cl}$
being the lifetime of molecular clouds, and $\tau$ being the typical
time-scale of grain growth. The typical time-scale of grain growth,
$\tau$, is given below in equation (\ref{eq:tau}). Note that
equation (\ref{eq:dmdt_acc}) is derived by assuming
that grain growth occurs in dense molecular
clouds which also host star formation. This is why the star formation
efficiency enters grain growth. Since the rate of grain growth
depends on the grain size distribution, $\beta$ depends on
the moments of grain radius as shown in equation (\ref{eq:beta}).  {The factor $(1-\xi)$ indicates that the dust mass increases with a larger fraction if a larger part of metals are in the gas phase, and thus the}  term, $(1-\xi )/\xi$, expresses the saturation grain growth for
$\xi\to 1$.

We adopt the following expression for $\tau$
 {\citep[Eq. 23, applicable for silicate]{hirashita11}}:
\begin{eqnarray}
\tau=6.3\times 10^7\left(\frac{Z}{\mathrm{Z}_{\sun}}\right)^{-1}
a_{0.1}n_3^{-1}T_{50}^{-1/2}S_{0.3}^{-1}~\mathrm{yr},\label{eq:tau}
\end{eqnarray}
where $a_{0.1}\equiv a_0/(0.1~\micron )$,
$n_3\equiv n_\mathrm{H}/(10^3~\mathrm{cm}^{-3})$
($n_\mathrm{H}$ is the number density of hydrogen nuclei),
$T_\mathrm{50}\equiv T_\mathrm{gas}/(50~\mathrm{K})$ ($T_\mathrm{gas}$
is the gas temperature), and $S_{0.3}\equiv S/0.3$ ($S$ is the
sticking probability of the dust-composing material onto the preexisting
grains.
We use the same values as in \citet{hirashita11}; i.e.\ $a_0 = 0.1~\micron$,
$n_\mathrm{H} = 10^3$ cm$^{-3}$, $T_\mathrm{gas} = 50$ K, and $S = 0.3$
unless otherwise stated.
 {A similar
time-scale is obtained for carbonaceous dust
mainly because silicate and carbonaceous dust have
similar total abundances of dust-composing materials}
 {\citep[Eq. 24]{hirashita11} Thus,}
we simply use equation (\ref{eq:tau}) for $\tau$ in this paper.

By using equation (\ref{eq:dmdt_acc}),
equation (\ref{eq:dDdZ}) can be restated as
\begin{eqnarray}
\mathcal{Y}\frac{\mathrm{d}\mathcal{D}}{\mathrm{d}Z}=f_\mathrm{in}
(\mathcal{R}Z+\mathcal{Y})-\left(\beta_\mathrm{SN}+\mathcal{R}-
\frac{\beta}{\epsilon}\right)\mathcal{D}.
\end{eqnarray}
We solve this equation to obtain the $\mathcal{D}-Z$ relation.

\section{Results}\label{sec:result}

\subsection{Overall behaviour of the $\mathcal{D}-Z$ relation} 
\label{subsec:theoretical}

\begin{figure*}
\includegraphics[width=0.45\textwidth]{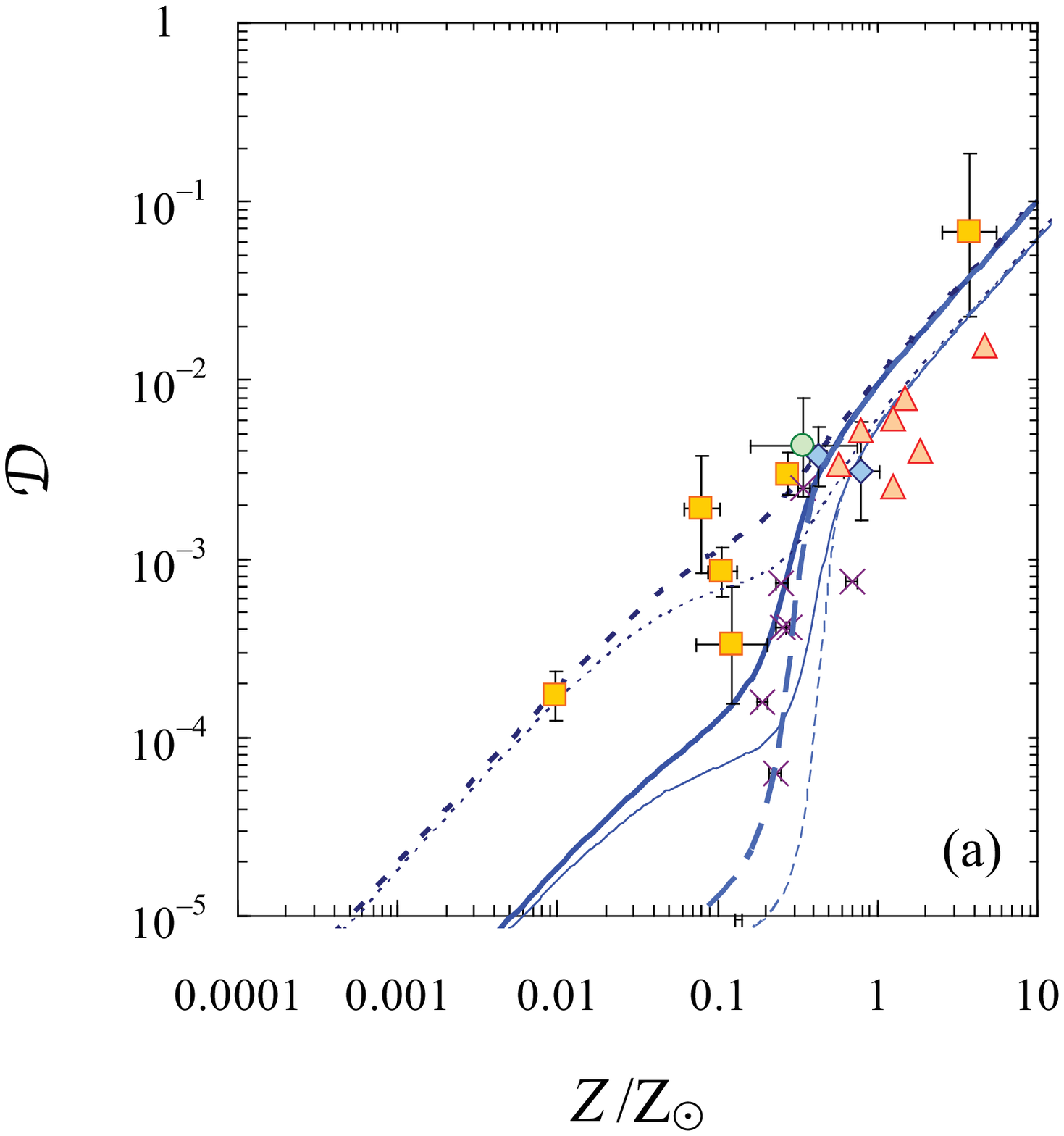}
\includegraphics[width=0.45\textwidth]{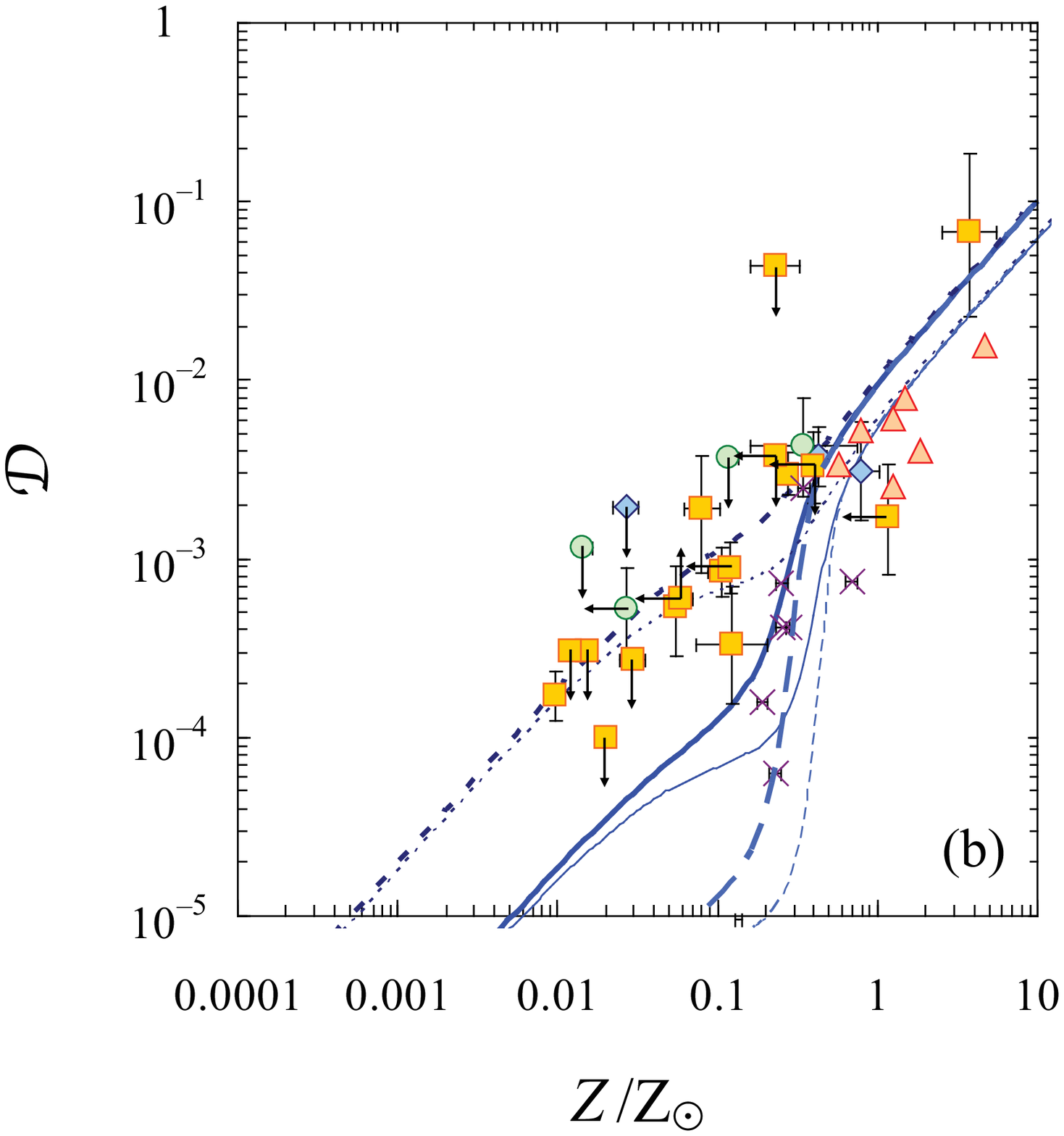}
\caption{
 Relation between the dust-to-gas ratio  $\mathcal{D}$ and
 metallicity $Z$. The squares, diamonds, and circles  {represent}
 the GRB sample data whose dust-to-gas ratios are derived by using
 SMC, LMC, and MW extinction curves, respectively \citep{zafar13}.
 The dashed, solid, and dotted lines indicate the evolution of
 dust content with $f_{\mathrm{in}}$ (the dust condensation efficiency
 of the metals in the stellar ejecta) equal to $0.01$, $0.1$, and $1$,
 respectively. The thick lines are for $\beta_{\mathrm{SN}}$ (the dust destruction factor by SN) equal to $9.65$, while the thin ones are with $\beta_{\mathrm{SN}} = 19.3$. For comparison, the nearby galaxy data are also shown, the crosses and triangles representing the blue compact dwarf galaxies and the spiral galaxies, respectively.  {Panel} (a) presents data points with all detected $A_V$, $N_{\mathrm{H}}$, and metallicity, while Panel (b) presents all the data points, with arrows indicating upper or lower limits.
\label{fig:dg_metal}}
\end{figure*}

The $\mathcal{D}-Z$ relations calculated by our models above
are shown for various values of $f_\mathrm{in}$ and $\beta_\mathrm{SN}$
in Fig.\ \ref{fig:dg_metal}.
The overall behaviour of the $\mathcal{D}-Z$ relation predicted
by the model has already been described in \citet{hirashita11}.
We briefly summarize it here.
In low metallicity environments, the dust is simply supplied from
the dust condensation in stellar ejecta, which results in an approximately
linear relation, $\mathcal{D}\sim f_\mathrm{in} Z$. Afterwards, when
the dust-to-gas ratio $\mathcal{D}$ reaches a value about
$f_\mathrm{in}\mathcal{Y} / \beta_\mathrm{SN}$, dust destruction
in SNe remnants suppresses the increase of the dust, and the
$\mathcal{D}-Z$ relation
becomes flatter at this stage. Ultimately, in high metallicity
environments, the dust-to-gas ratio $\mathcal{D}$ increases sharply
because of the nonlinear increase in dust content through
grain growth, i.e.\
$\mathrm{d}\mathcal{D} / \mathrm{d}Z\propto\mathcal{D}Z$.

\subsection{Comparison with GRB afterglows}
\label{subsec:comparison}

For GRBs, we first adopt merely 9 data points with both
detected $[A_V/N_\mathrm{H}]$ and metallicity. As shown in
Fig.\ \ref{fig:dg_metal}, it is obvious that some data
points at low metallicities can just be explained
with $f_\mathrm{in}\sim 1$, which indicates an extremely
efficient condensation of metals into dust in stellar ejecta.
The possibility of such a high $f_\mathrm{in}$ is further
discussed in terms of theoretical calculations of dust production
in stellar ejecta in the literature
(Section \ref{subsec:condensation}).

We also show the dependence on $\beta_\mathrm{SN}$ by
fixing $f_\mathrm{in}=1$. In Fig.\ \ref{fig:dg_metal}a, we show
the cases with $\beta_\mathrm{SN}=9.65$ (standard value) and
$19.3$ (two-times efficient dust destruction). We observe
that the difference appears around 0.01--0.1 Z$_{\sun}$;
since the dust destruction rate is proportional
to the dust-to-gas ratio, it is negligible compared with the dust
formation in stellar ejecta
at low metallicities ($\la 0.01$ Z$_{\sun}$) \citep{yamasawa11}.
At $Z>0.1$ Z$_{\sun}$, grain growth becomes efficient.
Therefore, the scatter of $\mathcal{D}$ around 0.01--0.1 Z$_{\sun}$
can be interpreted as different destruction efficiencies.
However, we should note that this interpretation is only
possible for $f_\mathrm{in}\sim 1$: if $f_\mathrm{in}\ll 1$, the
inclusion of efficient SN shock destruction makes it difficult to
explain the objects with relatively large $\mathcal{D}$
around $Z\sim 0.1$ Z$_{\sun}$.

Grain growth can raise $\mathcal{D}$. The effect of grain
growth appears in Fig.\ \ref{fig:dg_metal}a as
a rapid increase of $\mathcal{D}$ around $Z\sim 0.1$--1 Z$_{\sun}$.
Grain growth is prominent only for $f_\mathrm{in}\ll 1$, since
most of the metals are already in dust grains for $f_\mathrm{in}\sim 1$.
However, grain growth has difficulty in explaining the
high dust-to-gas ratios in low-metallicity objects since
grain growth becomes effective only when the ISM is
significantly enriched with metals \citep[e.g.][]{asano13}.

In Fig.\ \ref{fig:dg_metal}b, we also show the same results
including data with upper and lower limits. Two data around
$Z\sim 0.02$--0.03 Z$_{\sun}$ with upper limits for $\mathcal{D}$
can be explained by moderate condensation efficiencies with
$f_\mathrm{in}<1$, so there could be some variation in
$f_\mathrm{in}$. Other than those data points, the upper/lower
limit data do not constrain the model parameters more severely
than the 9 data points with detections. Thus, we hereafter
 {focus on the 9 data points} with detection in our analysis.

As shown in Fig.\ \ref{fig:dg_metal}, the lines with
$f_\mathrm{in}\sim 0.1$ is consistent with
the $\mathcal{D}-Z$ relation of the nearby sample.
The steep rise of $\mathcal{D}$ around
$Z\sim 0.1$ Z$_{\sun}$ is due to grain growth.
It is noteworthy that grain growth is
required to explain the steep trend of dust-to-gas ratio
relative to metallicity in
BCDs and the relatively high dust-to-gas ratio of
spiral galaxies. We also observe in Fig.\ \ref{fig:dg_metal}
that, at any metallicity, the dust-to-gas ratios for GRBs
are overall higher than those of BCDs or spiral galaxies.
Therefore, it appears that there is a
tension  {between the $f_\mathrm{in}$ value of GRBs
and that of nearby galaxies}.

\begin{figure}
\includegraphics[width=0.45\textwidth]{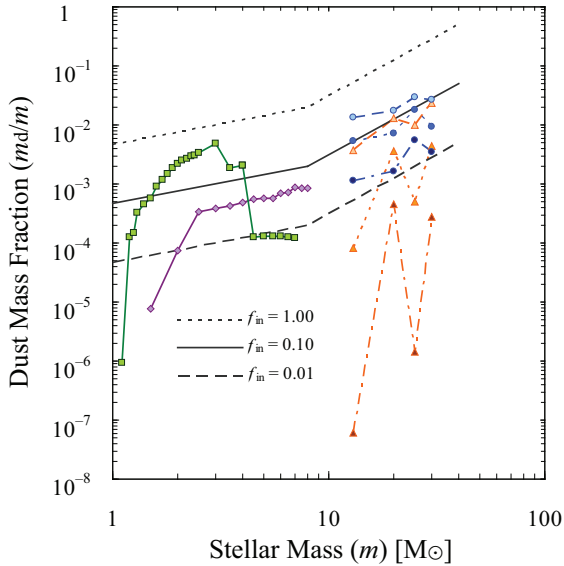}
\caption{
Ejected dust mass fraction $m_\mathrm{d}/m$ as a function of
the stellar mass at the zero-age main sequence. For AGB dust,
the data of the square
points are taken from \citet{zhukovska08} with metallicity $Z = 0.008$,
and the data of the diamond points are taken from
\citet{ventura12} with metallicity $Z = 0.008$. For SN dust,
the data of the round points are from top to bottom in sequence
with hydrogen density $n_\mathrm{H} = 0.1$, $1$, and $10$ cm$^{-3}$
for mixed helium cores, and the data of the triangular points are from top
to bottom in sequence with hydrogen density $n_\mathrm{H} = 0.1$,
$1$, and $10$ cm$^{-3}$ for unmixed helium cores, both taken from
\citet{nozawa07}. The three parallel black skew lines indicate
the double power-law approximations for the ejected dust mass fraction
rate provided by \citet{inoue11}, with condensation efficiencies
of metals into dust in stellar ejecta $f_\mathrm{in} = 1$,
$0.1$, and $0.01$, from the upper to the lower lines, respectively.
\label{fig:yield}}
\end{figure}

Summarizing the results, the GRB data can be explained with
$f_\mathrm{in}\sim 1$, which is significantly larger than the
value fitting the nearby galaxies ($f_\mathrm{in}\sim 0.1$).
In the next section, we discuss
(i) the value of $f_\mathrm{in}$ suggested by
dust condensation models in stellar ejecta;
(ii) alternative explanations; and (iii) possible observational
reasons for the tension between the GRB sample and the
nearby galaxy sample.

\section{Discussion}\label{sec:discussion}

\subsection{Condensation efficiency dust in stellar ejecta}
\label{subsec:condensation}

To discuss the reasonable range for the value of $f_\mathrm{in}$,
we refer to the data compiled by \citet{inoue11} for
theoretical calculations on dust formation in the wind of AGB
stars by \citet*{zhukovska08}, and on dust formation
in SN ejecta by \citet{nozawa07}. We add  {additional} data
for the dust formation in AGB stars calculated by
\citet{ventura12}. \citet{bianchi07}'s models
for the dust formation in SN ejecta are located in the range
consistent with \citet{nozawa07}'s results. For
\citet{nozawa07}'s models, we adopt ambient hydrogen
number densities of 0.1, 1, and 10 cm$^{-3}$ since more
dust is destroyed (so less dust is ejected) in high-density
environments. The relation between the progenitor stellar
mass $m$ at the zero-age main sequence and the dust mass
produced $m_\mathrm{d}$ is shown in Fig.\ \ref{fig:yield}
($m_\mathrm{d}$ is normalized to $m$). In order to constrain
$f_\mathrm{in}$ from the
plot, we need to assume the total metal mass ejected from
a star, $m_Z$. We follow \citet{inoue11} for the relation
between $m_Z/m$ and $m$:
\begin{eqnarray}
\frac{m_Z}{m}=\left\{
\begin{array}{ll}
0 & (m>40~\mathrm{M}_{\sun}),\\
0.02\, (m/8~\mathrm{M}_{\sun})^2 & (8~\mathrm{M}_{\sun}\leq
m\leq 40~\mathrm{M}_{\sun}),\\
0.02\, (m/8~\mathrm{M}_{\sun})^{0.7} & (m<8~\mathrm{M}_{\sun}).
\end{array}
\right.
\end{eqnarray}
Fig.\ \ref{fig:yield} shows the lines for $m_\mathrm{d}=f_\mathrm{in}m_Z$
for various $f_\mathrm{in}$.
We find that
$f_\mathrm{in}\sim 0.1$ is supported on average while
$f_\mathrm{in}=1$ is far above all the predictions.
Thus, such an extremely high condensation efficiency as
$f_\mathrm{in}\sim 1$ as suggested by the data points of
GRBs (Section \ref{subsec:comparison}) is somewhat
excessive. It is interesting that
$f_\mathrm{in}=0.1$ is consistent with the nearby galaxy
data in Fig.\ \ref{fig:dg_metal}.

\subsection{Grain growth}

\begin{figure*}
\includegraphics[width=0.45\textwidth]{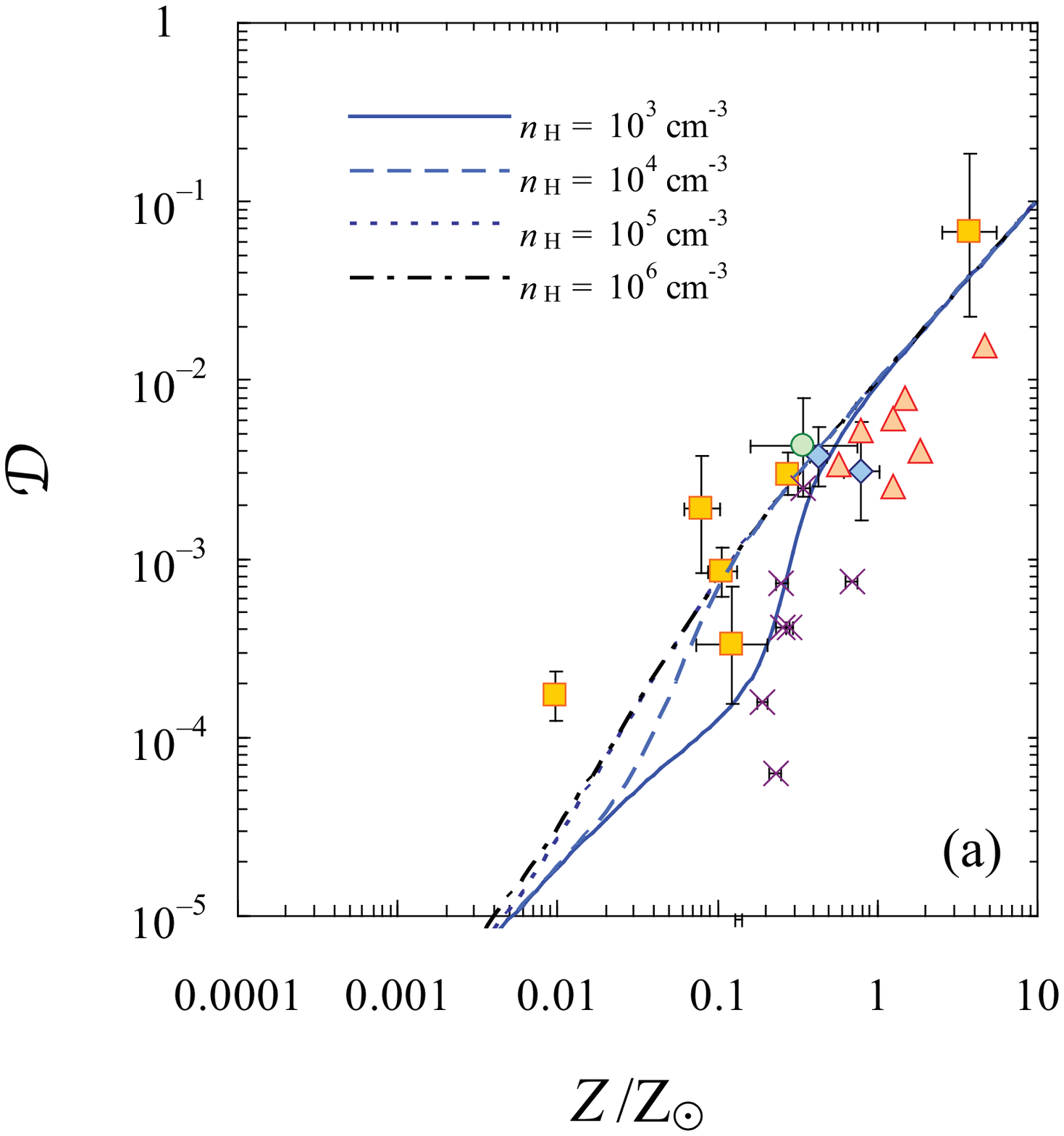}
\includegraphics[width=0.45\textwidth]{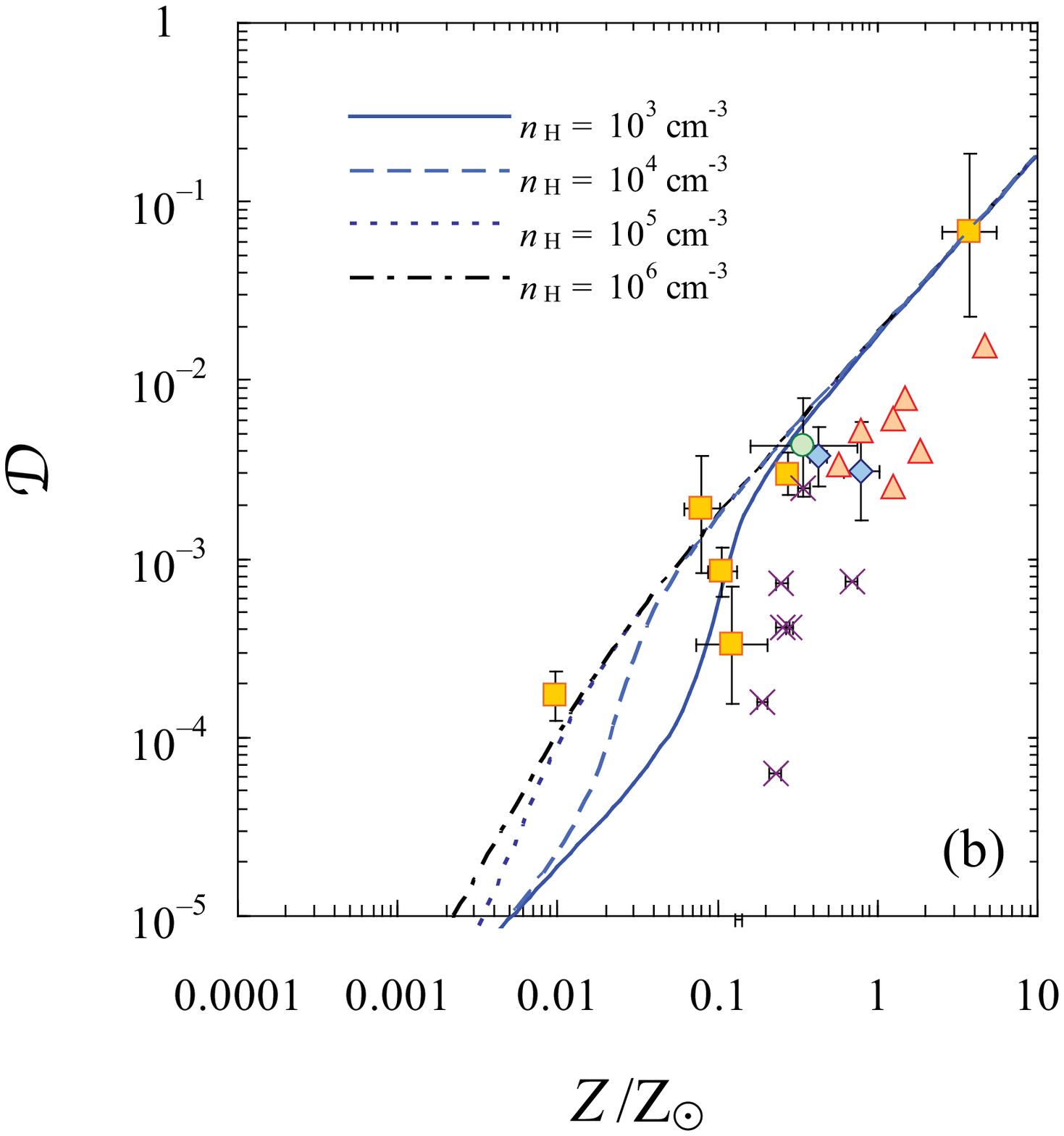}
\caption{
Same as Fig.\ \ref{fig:dg_metal}, but the solid, dashed, dotted,
and dot-dashed lines indicate the results with
$n_\mathrm{H}=10^3$,  $10^4$, $10^5$, and $10^6$ cm$^{-3}$,
respectively, and all with $f_{\mathrm{in}}=0.1$.
Panels (a) and (b) present the results with $\epsilon$ (the star
formation efficiency) $=0.1$ and 0.01, respectively.
\label{fig:dense}}
\end{figure*}

It is worth searching for an alternative solution that can explain
the high dust-to-gas ratios of the GRB sample. Here, we try
to explain the data with extremely efficient grain growth by
adopting a moderate $f_\mathrm{in}\sim 0.1$. An efficient
grain growth is equivalent to a short $\tau$, which can be
reasonably realized by a high $n_\mathrm{H}$  (equation \ref{eq:tau}).
In Fig.\ \ref{fig:dense}a, we show the $\mathcal{D}-Z$ relation for
$n_\mathrm{H}=10^3$, $10^4$, $10^5$, and $10^6$ cm$^{-3}$.
The enhanced grain growth actually explain the data points around
$Z=0.1$ Z$_{\sun}$; however, the excessively large dust content
of some data points at low metallicities ($Z<0.1$ Z$_{\sun}$) is still
not reproduced. This is because the grain growth time-scale is typically
longer than the metal enrichment time-scale at such low metallicities
as estimated below.

Let us compare the metal enrichment time-scale and the
grain growth time-scale. After a significant gas consumption,
the metallicity becomes roughly solar \citep[e.g.][]{tinsley80}.
Therefore, the
metal enrichment time-scale to a metallicity level of $Z$ is
estimated as
$\tau_Z\sim (M_\mathrm{gas}/\psi)(Z/\mathrm{Z}_{\sun})$,
where $M_\mathrm{gas}$ is the gas mass.
The grain growth time-scale is, on the other hand,
$\tau_\mathrm{grow}\sim M_\mathrm{dust}/
[\mathrm{d}M_\mathrm{dust}/\mathrm{d}t]_\mathrm{acc}
=\epsilon M_\mathrm{gas}/(\beta\psi )=
\epsilon \tau_\mathrm{Z}(\mathrm{Z}_{\sun}/Z)/\beta$.
At the metallicity level where grain growth becomes the
dominant dust-producing mechanism, $\beta\sim 1$
\citep{hirashita11}.
Thus, $\tau_Z<\tau_\mathrm{grow}$ for $Z/\mathrm{Z}_{\sun}<\epsilon$,
which indicates that the metal enrichment occurs faster than the
grain growth at $Z<0.1$ Z$_{\sun}$ if $\epsilon =0.1$ as adopted
above. In other words, grain growth cannot
affect the dust-to-gas ratio by the time when the system is
enriched with metals up to 0.1 Z$_{\sun}$, as long as we adopt
$\epsilon =0.1$ derived from nearby molecular clouds
\citep[e.g.][]{lada10}.

The above arguments suggest that grain growth becomes efficient
at even lower metallicities if we adopt smaller $\epsilon$.
In Fig.~\ref{fig:dense}b, we show the results for $\epsilon =0.01$.
We observe that, considering denser gas and less efficient star
formation, we can reproduce the data points at $Z<0.1$ Z$_{\sun}$.
This is because low star formation efficiency allows grains
to grow within the metal enrichment time-scale,
which is inversely proportional to the star formation efficiency.
Thus, it appears that the only way to explain the dust-to-gas ratio
in the GRB sample at $Z<0.1$ Z$_{\sun}$ is to impose
a low star formation efficiency (or equivalently to slow
the metal enrichment). This implies the difference
between nearby galaxies and the host galaxies of GRBs in
terms of the star formation efficiency in dense clouds, although
the physical mechanism of producing such a difference is
not clear. Alternatively, the GRB sample is biased to objects
with low star formation efficiencies, as discussed in the next
subsection.

In summary, slow metal enrichment because of
low star formation efficiency enable grain growth to occur
within the metal enrichment time-scale even at low metallicities,
so that most of
the metals can be condensed into dust by grain growth.
Therefore, the combination of a low star formation
efficiency and an enhanced grain growth by a high density
can explain the data points at low metallicities.

\subsection{Comparison with the nearby galaxy sample}

As shown in Fig.\ \ref{fig:dg_metal}, the dust-to-gas ratios in
GRBs are systematically larger than those of nearby galaxies.
According to the argument in the previous subsection,
a possible interpretation is that GRB host galaxies have smaller
star formation efficiencies than nearby galaxies. Probably,
the nearby galaxy sample is biased to star-forming objects,
which are by definition forming stars actively. On the other
hand, we tend to choose gas-rich galaxies by a GRB, since
we sample objects whose absorption by the ISM in
the host galaxy is significantly detected. Therefore, GRBs also
potentially include quiescent galaxies whose star formation efficiency
is not large.

\citet{zafar13} point out that in estimating the dust-to-gas ratio
in nearby dwarf galaxies, we have to take into account
the fact that the entire H \textsc{i} gas is
clearly more extended than the dust emission.
Indeed, \citet{draine07} show that, if we consider only
the regions over which the dust emission is detected,
a substantial fraction of interstellar dust-composing
materials appears to be contained in dust. The dust mass
estimate itself may also be changed: \citet{galametz11}
show that the addition of submillimetre data results in
higher dust masses than without submillimetre data.
These two effects tend to
make the difference in the dust-to-gas ratio
between the nearby galaxy sample and the GRB sample
smaller.

It is worth noting that the dust-to-metals ratios of the GRB sample
is almost the same as that of the Milky Way, as pointed out by
\citet{zafar13}. \citet{mattsson13} have recently shown by using
an analytic model that such a `constant dust-to-metals ratio'
can be understood as the result of a balance between destruction
and growth of grains in the ISM.  In particular, they show that,
if we consider a certain balance between grain growth and destruction,
a constant dust-to-metals ratio may naturally be obtained as
an asymptotic solution; i.e.\ the system tends to converge to
a constant dust-to-metals ratio. This suggests a `unified model'
may be constructed, which can explain both the GRB sample and
other data without a lot of parametric fine tuning. However, the importance of grain growth is a common feature between our model and theirs, although we did not focus on dust destruction.

\section{Conclusion}\label{sec:conclusion}

The dust-to-gas ratios derived from GRB afterglow
extinction data are excessively high such that they can be explained
with an extremely efficient condensation of metals into dust
in stellar ejecta, while theoretical calculations on dust formation
in the wind of AGB stars and in the ejecta of SNe suggest much
more moderate condensation efficiencies. We alternatively adopt
a moderate condensation efficiency and a more
efficient grain growth in dense clouds. Even with efficient
grain growth, the excessive dust-to-gas ratio can
only be explained if we assume a low star formation efficiency,
which is equivalent with slow metal enrichment.
Therefore, some GRB host galaxies in which
most of the dust-composing metals are condensed into dust
can be explained  with enhanced grain growth in dense clouds,
whose time-scale
should be shorter than the metal enrichment time-scale
(or equivalently the star formation efficiency in dense clouds
is as small as $\la 0.01$).

\section*{Acknowledgments}
We are grateful to A. K. Inoue for providing us with a
compiled data set of dust formation in stellar sources.
We also thank L. Mattsson, D. Watson, and A. C. Andersen
for helpful discussions.
This research is supported through NSC grant 99-2112-M-001-006-MY3.



\bsp

\label{lastpage}

\end{document}